%% file: ITW10.tex
\documentclass[conference]{IEEEtran}
\usepackage{cite}

\usepackage{graphicx}
\usepackage{epsfig}
\usepackage{subfigure}

\usepackage[cmex10]{amsmath}
\usepackage{amssymb}
\usepackage{wasysym}
\usepackage{algorithmic}

\usepackage{array}

\usepackage{url}

\usepackage{color}

\newtheorem{te}{Theorem}

\newtheorem{ex}{Example}
\newtheorem{prop}{Proposition}

\begin{document}

\input{Title_v2}

\input{Abstract_v5}

\input{Introduction_v7}

\input{Preliminaries_v5}

\input{GeneralConstruction_v1}

\input{ConstructionAdditive_v3}


\input{ProgressiveConstruction_v5}

\input{Conclusion_v2}

\input{Acknowledgment}
\bibliographystyle{IEEEtran}
\bibliography{reference}

\end{document}

%% file: Title_v2.tex
%
\title{Structured LDPC Codes from Permutation Matrices Free of Small Trapping Sets}

\author{\IEEEauthorblockN{Dung Viet Nguyen, Bane Vasi$\acute{\mathrm{c}}$, Michael Marcellin}
\IEEEauthorblockA{Department of Electrical and Computer Engineering\\
University of Arizona, Tucson, AZ 85721\\
Email: \{nguyendv, vasic, marcellin\}@ece.arizona.edu}
\and
\IEEEauthorblockN{Shashi Kiran Chilappagari}
\IEEEauthorblockA{Marvell Semiconductor Inc, \\
5488 Marvell Lane, Santa Clara, CA 95054\\
Email: shashickiran@gmail.com}}


%


\maketitle

%% file: Abstract_v5.tex
\begin{abstract}
This paper introduces a class of structured low-density parity-check (LDPC) codes whose parity check matrices are arrays of permutation matrices. The permutation matrices are obtained from Latin squares and form a finite field under some matrix operations. They are chosen so that the Tanner graphs do not contain subgraphs harmful to iterative decoding algorithms. The construction of column-weight-three codes is presented. Although the codes are optimized for the Gallager A/B algorithm over the binary symmetric channel (BSC), their error performance is very good on the additive white Gaussian noise channel (AWGNC) as well.
\end{abstract}

%% file: Introduction_v7.tex
\section{Introduction}
It is now well established that the error floor phenomenon, an abrupt degradation in the error performance of low-density parity-check (LDPC) codes \cite{ldpcBook_gallager}  in the high signal-to-noise-ratio (SNR) region, is due to the presence of certain structures in the Tanner graph that lead to decoder failures. For iterative decoding, these structures are known as trapping sets.

Ideally, LDPC codes should be designed so that their Tanner graphs do not contain most harmful trapping sets, but unfortunately, except for the binary erasure channel, trapping sets for other channels such as the BSC or the AWGNC, are only partially understood. Consequently, many existing methods of constructing LDPC codes only attempt to maximize the girth of the Tanner graphs \cite{shortenedArrayCode_milenkovic,highGirthQC_yedidia,qcGirth14_koreanGuy} or avoid subgraphs that are believed to be harmful \cite{07DZAWN}. In the latter approach, the subgraphs are identified either by computer simulation or hardware emulation, or are conveniently defined to make the search easier. The problem in these approaches lies in the underlying assumption about harmfulness, which is not proven or is restricted to specific cases.

In this paper, LDPC codes are constructed so that their Tanner graphs do not contain trapping sets of the Gallager A/B algorithm on the BSC. The code construction utilizes the \textit{Trapping Set Ontology} (TSO) given by Vasic \textit{et. al.} in \cite{ontology_vasic}. This database contains trapping sets for the Gallager A/B algorithms which are organized based on their topological relations. Our approach relies greatly on the \textit{relative harmfulness}\footnote{The relative harmfulness of a trapping set given in the TSO for the Gallager A/B algorithm over the BSC is determined by its critical number and strength \cite{errorFloorBSC_shashi}. The relative harmfulness of a trapping set given in the TSO for the sum product algorithm over the BSC is currently being studied and will be discussed in the journal version of this paper.} of different trapping sets to determine which trapping sets should not be present in the Tanner graphs of codes. The choice of which trapping sets to avoid is critical to the error performance and the code rate.

Although the codes are optimized for the Gallager A/B algorithm over the BSC, experimental results indicate that their error performance under other iterative algorithms such as the sum product algorithm (SPA) are also extremely good. The explanation for this is based on the observation by Chilappagari \textit{et. al.} in \cite{instantonPaper_shashi} that the decoding failures for various decoding algorithms and channels are closely related and subgraphs responsible for these failures share some common underlying topological structures. These structures are either trapping sets for iterative decoding algorithms on the BSC or larger subgraphs containing these trapping sets.

The above approach can be incorporated into many existing classes of LDPC codes to result in codes with good error performance (e.g., \cite{qcFiniteField_lanLin,arrayCode_fan}), including the new class of structured LDPC codes that we propose in this paper.

To be efficiently encodable  and decodable, LDPC codes must be structured. An important class of structured LDPC codes are quasi-cyclic (QC) codes. In the past few years, numerous QC constructions have been proposed. They can be broadly classified as algebraic \cite{tannerCode_tanner,reedSol_ivana,nearShanonCode_ivana,qcFiniteField_lanLin} and combinatorial \cite{finiteGeometryLDPC_kouLin, combinatorialLDPC_vasic}. In this paper, we give a class of structured LDPC codes whose parity check matrices are arrays of permutation matrices. Our design is motivated by the work of Lan \textit{et. al.} \cite{qcFiniteField_lanLin}. In \cite{qcFiniteField_lanLin}, the authors give a general algebraic construction of QC-LDPC codes based on a one-to-one correspondence between an element of the multiplicative group of GF($q$) and a circulant permutation matrix of size $(q-1)\times (q-1)$. In our construction, the set of permutation matrices together with some matrix operations (introduced later in this paper) form a field isomorphic to GF($q$). Our permutation matrices are similar to circulants in the sense that the set of $q-1$ non-identity permutation matrices form a cyclic group, but they are more general as the circulant property holds on indices understood as elements of GF($q$). More specifically, the permutation corresponding to $\alpha^t$ sends the indices $(0,1,\alpha,\ldots,\alpha^{q-2})$ to $(0+\alpha^t,1+\alpha^t,\alpha+\alpha^t,\ldots,\alpha^{q-2}+\alpha^t)$.

The construction allows for a systematic reduction of error floors. We present a construction algorithm which recursively builds the parity check matrix by adding permutation matrices. The algorithm ensures that after each step, the corresponding Tanner graph does not contain certain trapping sets defined in the TSO.

The rest of the paper is organized as follows. In Section \ref{sec_preliminaries}, we provide the background related to LDPC codes and the necessary preliminaries for the construction method. In Section \ref{generalConstruction}, we give the general definition of the class of structured LDPC codes from permutation matrices. In Section \ref{sec_additive}, we present the construction of codes based on Latin squares obtained from the additive group of a Galois field. In Section \ref{progressiveConstruction}, we describe the construction algorithm and present the construction of several column-weight-three codes. Finally, we conclude the paper in Section \ref{discussion}.

%% file: Preliminaries_v5.tex
\section{Preliminaries}\label{sec_preliminaries}
In this section, we introduce the definitions and notation used throughout the paper.
\subsection{LDPC Codes and Trapping Sets}
Let $\mathcal{C}$ denote an ($n,k$) LDPC code over the binary field GF(2). $\mathcal{C}$ is defined by the null space of $H$, an $m\times n$ \textit{parity check matrix} of $\mathcal{C}$. $H$ is the bi-adjacency matrix of $G$, a Tanner graph representation of $\mathcal{C}$. $G$ is a bipartite graph with two sets of nodes: $n$ variable (bit)
nodes $V = \{1, 2,\ldots, n\}$ and $m$ check (constraint) nodes $C = \{1, 2,\ldots ,m\}$. The length of the shortest cycle in the Tanner graph $G$ is called the girth $g$ of $G$.

A trapping set for an iterative decoding algorithm is defined as a non-empty set of variable nodes in $G$ that are not eventually corrected by the decoder \cite{errorFloor_richarson}. A trapping set $\mathcal{T}$ is called an ($a,b$) trapping set if it contains $a$ variable nodes and the subgraph induced by these variable nodes has $b$ odd degree check nodes.
%
\subsection{Permutation Matrices from Latin Squares}
A permutation matrix is a square binary matrix that has exactly one entry 1 in each row and each column and 0's elsewhere. Our construction makes use of permutation matrices that do not have 1's in common positions. These sets of permutation matrices can be obtained conveniently from Latin squares.

A \textit{Latin square} of \textit{size} $q$ (or \textit{order} $q$) is a $q\times q$ array in which each cell contains a single symbol from a $q$-set $S$, such that each symbol occurs exactly once in each row and exactly once in each column. A Latin square of size $q$ is equivalent to the \textit{Cayley table} of a quasigroup $\mathcal{Q}$ on $q$ elements (see \cite[pp. 135--152]{comDesignBook} for details).

For mathematical convenience, we use elements of $\mathcal{Q}$ to index the rows and columns of Latin squares and permutation matrices. Let $\mathcal{L}={[l_{i,j}]}_{i,j\in \mathcal{Q}}$ denote the Latin square defined on the Cayley table of a quasigroup ($\mathcal{Q},\oplus$) of order $q$. Define $f$, an injective map from $\mathcal{Q}$ to $\mbox{Mat}(q,q,\mbox{GF}(2))$, the set of matrices of size $q\times q$ over GF(2), as follows:
\begin{eqnarray}\label{eq_Fdef}
f: \mathcal{Q}&\rightarrow& \mbox{Mat}(q,q,\mbox{GF}(2))\nonumber\\
\alpha &\mapsto& f(\alpha)={[m_{i,j}]}_{i,j\in \mathcal{Q}}\nonumber
\end{eqnarray}
such that:
\begin{eqnarray}
m_{i,j} &=& \left\{\begin{array}{cc}
1 &\mbox{~if~} l_{i,j} = \alpha\\
0 &\mbox{~if~} l_{i,j} \neq \alpha
\end{array}\right..\nonumber
\end{eqnarray}

It follows from the above definition that the images of elements of $\mathcal{Q}$ under $f$ give a set of $q$ permutation matrices that do not have 1's in common positions.
%

%% file: GeneralConstruction_v1.tex
\section{LDPC Codes as Array of Permutation Matrices}\label{generalConstruction}
In this section, we give the general definition of LDPC codes whose parity check matrices are arrays of permutation matrices.
Let $\mathcal{W} = {[w_{i,j}]}_{1\leq i\leq \mu,1\leq j\leq \eta}$ be an $\mu\times \eta$ matrix over a quasigroup $\mathcal{Q}$,
\begin{eqnarray}\label{eq_Adef}
\mathcal{W} = \left[\begin{array}{cccc}
w_{1,1}&w_{1,2}&\cdots&w_{1,\eta}\\
w_{2,1}&w_{1,2}&\cdots&w_{2,\eta}\\
\vdots&\vdots&\ddots&\vdots\\
w_{\mu,1}&w_{\mu,2}&\cdots&w_{\mu,\eta}
\end{array}
\right].
\end{eqnarray}

With some abuse of notation, let $\mathcal{H} = f(\mathcal{W}) = {[f(w_{i,j})]}$ be an array of permutation matrices, obtained by replacing elements of $\mathcal{W}$ with their images under $f$, i.e., 
\begin{eqnarray}\label{eq_Hdef}
\mathcal{H} = \left[\begin{array}{cccc}
f(w_{1,1})&f(w_{1,2})&\cdots&f(w_{1,\eta})\\
f(w_{2,1})&f(w_{1,2})&\cdots&f(w_{2,\eta})\\
\vdots&\vdots&\ddots&\vdots\\
f(w_{\mu,1})&f(w_{\mu,2})&\cdots&f(w_{\mu,\eta})\\
\end{array}
\right].
\end{eqnarray}
Then $\mathcal{H}$ is a binary matrix of size $\mu q\times \eta q$. The null space of $\mathcal{H}$ gives an LDPC code of length $\eta q$. The column weight and row weight of $\mathcal{H}$ are $\mu$ and $\eta$, respectively.

%% file: ConstructionAdditive_v3.tex
\section{Structured LDPC Codes from Galois Fields of Permutation Matrices}\label{sec_additive}
\subsection{Galois Fields of Permutation Matrices}

Consider the Galois field GF($q$), where $q$ is a power of a prime. Let $\alpha$ be a primitive element of GF($q$). The powers of $\alpha$, $\alpha^{-\infty}\triangleq 0, \alpha^{0}= 1, \alpha, \alpha^{2},\ldots,\alpha^{q-2}$, give all the $q$ elements of GF($q$) and $\alpha^{q-1} = 1$. Let $\mathcal{L}={[l_{i,j}]}_{i,j\in \mbox{GF}(q)}$ denote the Latin square defined by the Cayley table of the quasigroup given by the set $\{0,1,\alpha,\ldots,\alpha^{q-2}\}$ together with the subtractive operation of GF($q$), i.e., $l_{i,j} = i-j$. Let $\mathcal{M} = \{M_{-\infty}, M_0, M_1, \ldots, M_{q-2}\}$ be the set of images of elements of GF($q$) under $f$, i.e., $M_t = {[m^{(t)}_{i,j}]}_{i,j\in \mbox{GF}(q)} = f(\alpha^t)$. It is easy to see that $M_{-\infty}=I$, the $q\times q$ identity matrix. To show that $\mathcal{M}$ forms a field isomorphic to GF($q$) under the matrix operations defined below, we give the following propositions. Due to page limitations, the proofs are omitted.
\begin{prop}\label{prop_addi}
For all $t_1, t_2 \in\mathbb{Z}$, $f(\alpha^{t_1}+\alpha^{t_2}) = M_{t_1}M_{t_2}$.
\end{prop}
\begin{prop}\label{pmq_prop}\label{prop_multi}
For all $t\geq0$, $M_{t+1} = PM_{t}Q$, where $P\in \mathcal{M}_{q\times q}[\mbox{GF}(2)]$ is given as
\begin{eqnarray}
P = \left[\begin{array}{cccccc}
1&0&0&\cdots&0&0\\
0&0&0&\cdots&0&1\\
0&1&0&\cdots&0&0\\
0&0&1&\cdots&0&0\\
\vdots&\vdots&\vdots&\ddots&\vdots&\vdots\\
0&0&0&\cdots&1&0\\
\end{array}
\right],
\end{eqnarray}
and $Q$ is the transpose of $P$.
\end{prop}

Define the addition $\boxplus$ and the multiplication $\boxdot$ on $\mathcal{M}$ as:
\begin{eqnarray}
M_{t_1}\boxplus M_{t_2} &=& M_{t_1}M_{t_2},\nonumber\\
M_{t_1}\boxdot M_{t_2} &=& P^{(t_2-t_1)}M_{t_1}Q^{(t_2-t_1)}\nonumber\\
&=& P^{(t_1-t_2)}M_{t_2}Q^{(t_1-t_2)}\nonumber
\end{eqnarray}
then it can be shown that $\mathcal{M}$ together with $\boxplus$ and $\boxdot$ form a field isomorphic to GF($q$).
\subsection{LDPC Codes from Galois Fields of Permutation Matrices}
Define $\mathcal{W}$ and $\mathcal{H}$ as in (\ref{eq_Adef}) and (\ref{eq_Hdef}), where $\mathcal{Q}$ is the set $\{0,1,\alpha,\ldots,\alpha^{q-2}\}$ together with the subtractive operation of GF($q$). The following theorem gives the necessary and sufficient condition on $\mathcal{\mathcal{W}}$, such that the Tanner graph corresponding to $\mathcal{H}$ has girth at least 6.
\begin{te}[Cross-addition Constraint]\label{te_crossAddi}
The Tanner graph corresponding to $\mathcal{H}$ contains no cycle of length four iff $w_{i_1,j_1}+w_{i_2,j_2}\neq w_{i_1,j_2}+w_{i_2,j_1}$ for any $1\leq i_1,i_2 \leq \mu,1\leq j_1,j_2 \leq \eta$, $i_1\neq i_2, j_1\neq j_2$.
\end{te}
\proof The proof is omitted due to page limitations. \endproof
It can be seen that the construction of LDPC codes with girth at least 6 from a Galois field of permutation matrices reduces to the finding of  a matrix $\mathcal{W}$ that satisfies the cross-addition constraint. One form of $\mathcal{W}$ that satisfies the cross-addition constraint is given by
\begin{eqnarray}\label{eq_Wdef}
\mathcal{W} = \left[\begin{array}{ccccc}
0&0&0&\!\cdots&0\\
0&1&\alpha&\cdots&\alpha^{q-2}\\
0&\alpha&\alpha^2&\cdots&1\\
\vdots&\vdots&\vdots&\ddots&\vdots\\
0&\alpha^{q-2}&1&\cdots&\alpha^{q-3}\\
\end{array}\right].
\end{eqnarray}

Let $\mathcal{H} = f(\mathcal{W})$. From Proposition \ref{pmq_prop}, it follows that $\mathcal{H}$ has the following structure:
\begin{eqnarray}\label{eq_Hdef2}
\mathcal{H} = \left[
\begin{array}{ccccc}
I&I&I&\cdots&I\\
I&M_0&M_1&\cdots&M_{q-2}\\
I&M_1&M_2&\cdots&M_0\\
\vdots\!\!&\vdots&\vdots&\ddots&\vdots\\
I&M_{q-2}&M_0&\cdots&M_{q-3}\\
\end{array}
\right],
\end{eqnarray}
where $M_t = P^tM_0Q^t$ and $I$ is the $q\times q$ identity matrix. $\mathcal{H}$ is an array of permutation matrices from $\mathcal{M}$ and is a $q^2\times q^2$ matrix over GF($q$) with both row and column weights $q$. Since $\mathcal{W}$ satisfies the cross-addition constraint, the Tanner graph corresponding to $\mathcal{H}$ contains no cycle of length 4.

For any pair ($\gamma,\rho$) of positive integers with $1\leq \gamma,\rho \leq q$, let $H$ be a $\gamma\times\rho$ subarray of $\mathcal{H}$. Then $H$ is a $\gamma q\times \rho q$ matrix over GF(2) which is also free of cycles of length 4. $H$ has constant column $\gamma$ and row weight $\rho$. The null space of $H$ gives a regular structured LDPC code $\mathcal{C}$ of length $\rho q$ with rate at least $(\rho-\gamma)/\rho$ \cite{ldpcBook_gallager}.

\textit{Remarks}: 
\begin{itemize}
\item The matrix $\mathcal{W}$ in (\ref{eq_Wdef}) is obtained by adding a row and a column of all zeros to $\mathcal{L}$, where $\mathcal{L}$ is the Latin square obtained from the Cayley table of the multiplicative group of $\mbox{GF}(q)$.
\item The codes given in this paper can be alternatively defined on integer latices. Since array LDPC codes introduced by Fan in \cite{arrayCode_fan}, can be defined on integer lattices as shown in \cite{highRateG8_vasic}, they are special cases of the codes given in this paper. If $q$ is a prime, then the parity check matrices of array LDPC codes are subarrays of $f(\mathcal{W}_p)$, where $\mathcal{W}_p$ is obtained by permuting rows and columns of $\mathcal{W}$ (in (\ref{eq_Wdef})). The codes by Lan \textit{et. al.} \cite{qcFiniteField_lanLin} based on the additive groups of prime fields are also array LDPC codes.
\item Our class of codes is also different from codes given by Gabidulin \textit{et. al} in \cite{generalPermCode_gabidulin} (except for codes based on prime fields, for which the latter become array codes). In \cite{generalPermCode_gabidulin}, permutation matrices of size $q\times q$, where $q$ is a power of a prime, are obtained from the Tensor product of circulant matrices of size $p\times p$, thus are different from the permutation matrices in (\ref{eq_Hdef2}).
\item If $\mathcal{L}$ is defined by the Cayley table of the multiplicative group of GF($q$), then circulant permutation matrices of size $(q-1)\times(q-1)$ are obtained as images of elements of GF($q$)$\backslash\{0\}$. In such case, the necessary and sufficient condition on $\mathcal{W}$ such that the Tanner graph corresponding to $\mathcal{H}$ has girth at least 6 is called the \textit{cross-multiplication constraint}. This condition can be obtained from the cross-addition constraint by replacing addition with multiplication. This gives an alternative description for the codes described in \cite{qcFiniteField_lanLin}.
\end{itemize}

%% file: ProgressiveConstruction_v5.tex
\section{Construction of Codes Free of Small Trapping Sets}\label{progressiveConstruction}
\begin{figure}
\centering
\subfigure[] 
{
    \label{531TS}

\includegraphics[height = 1.1 in]{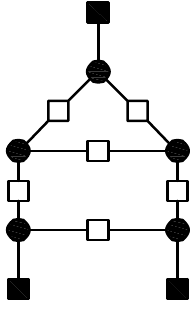}
}
\hspace{0.2in}
\subfigure[] 
{
    \label{532TS}

\includegraphics[height = 1.1 in]{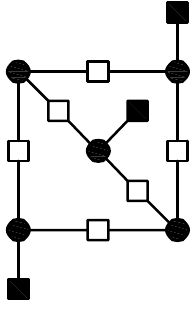}
}
\hspace{0.2in}
\subfigure[] 
{
    \label{643TS}

\includegraphics[height = 1.1 in]{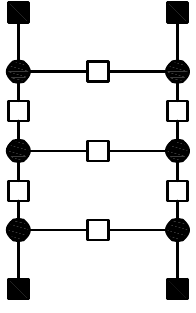}
}
\caption{\subref{531TS} ($5,3$) trapping set of girth 6, \subref{532TS} ($5,3$) trapping set of girth 8 and \subref{643TS} ($6,4$) trapping set. We use $\CIRCLE$ to represent variable nodes, $\blacksquare$ to represent odd degree check nodes and $\Box$ to represent even degree check nodes.}
\label{TS}
\end{figure}

The description of the class of LDPC codes given in the previous sections along with Theorem \ref{te_crossAddi} allow us to construct codes by progressively building the Tanner graphs. The construction is performed by an algorithm which forms the matrix $\mathcal{W}$ in (\ref{eq_Adef}). The algorithm is based on a check and select-or-disregard procedure. Let $\tau$ specify which graphical structures should not be present in the Tanner graph $G$. For example, Figure \ref{TS} shows the subgraphs induced by some small trapping sets. $\tau$ may specify the girth of $G$ and may also specify the minimum distance of the code. For column-weight-three codes, all possible codewords of even weight less than 12 are known and their induced subgraphs are listed in the TSO. It is simple to check the Tanner graph for cycles of length four thanks to Theorem \ref{te_crossAddi}. Finding girth of the Tanner graph can be done in polynomial time using the Dijkstra or Bellman-Ford algorithm, while enumerating cycles of a given length using a standard tree-based algorithm has linear complexity in the code length \cite{ontology_vasic}.  An efficient search of the Tanner graph for trapping sets relies on the topological relations among them and carefully analyzing the induced subgraphs. Details on the graph searching techniques will be given in the journal version of this paper.

The Tanner graph of a code is built progressively in $\rho$ stages, where $\rho$ is the row weight of the parity check matrix. Usually, $\rho$ is not pre-specified, and codes are constructed to have rate as high as possible. At each stage, a set of $q$ variable nodes are introduced, initially not connected to check nodes of the Tanner graph. Blocks of edges are then added to connect the new variable nodes and the check nodes. Each block of edges corresponds to a permutation matrix and hence corresponds to an element of GF($q$). An element of GF($q$) may be chosen randomly, or it may be chosen in a predetermined order. After a block of edges is added, the Tanner graph is checked for condition $\tau$. If the condition $\tau$ is violated, then that block of edges is removed and replaced by a different block. The algorithm proceeds until no block of edges can be added without violating condition $\tau$. It can be seen that the algorithm is a combination of the progressive edge growth algorithm for constructing random LDPC codes \cite{peg_hu} and the method in \cite{highRateG8_vasic}.




The complexity of the algorithm grows exponentially with the column weight. The complexity also depends greatly on how condition $\tau$ is checked on a Tanner graph. However, for small column weights, say 3 or 4, and small to moderate code lengths, the algorithm is well handled by state-of-the-art computers. For example, the construction of a ($1111,808$) code which has girth 8, minimum distance at least 10 and which contains no ($5,3$) trapping set given in Figure \ref{TS}\subref{532TS} takes less than 2 minutes on a 2.4 GHz computer.

We continue this section by providing two examples of construction of column-weight-three codes whose Tanner graphs do not contain small trapping sets described in the TSO.
\begin{ex}\label{ex_530}
Let $q=53$ and let $\mathcal{C}^{(\mathrm{new})}_8$ denote the code obtained when $\tau$ is defined as a condition that the corresponding Tanner graph of $\mathcal{C}^{(\mathrm{new})}_8$ has girth $g\geq 8$. $\mathcal{C}^{(\mathrm{new})}_8$ is a (530, 371) LDPC code with rate $R=0.7$. Let $\mathcal{C}^{(\mathrm{old})}_8$ denote the (530, 371) shortened array code (or integer lattice code described in \cite{highRateG8_vasic}). $\mathcal{C}^{(\mathrm{old})}_8$ is obtained by extensive computer search and has the maximum possible rate of $R=0.7$. Denote by $\mathcal{C}_{\mbox{d}10}$ the code obtained when $\tau$ is such that the minimum distance of $\mathcal{C}_{\mbox{d}10}$ is at least 10. $\mathcal{C}_{\mbox{d}10}$ is constructed by avoiding codewords of weight 6 and 8 in the Tanner graph (the TSO lists two possible codewords of weight 6 and five possible codewords of weight 8 for codes with $g\geq 6$). $\mathcal{C}_{\mbox{d}10}$ is a (795, 636) LDPC code with rate $R=0.8$ and girth $g=6$. The error performance of $\mathcal{C}^{(\mathrm{new})}_8$, $\mathcal{C}^{(\mathrm{old})}_8$ and $\mathcal{C}_{\mbox{d}10}$ under the SPA with a maximum of 50 iterations over the AWGNC is shown in Figure \ref{fg_503g8}. It can be seen that the error performance of $\mathcal{C}^{(\mathrm{new})}_8$ is better than that of $\mathcal{C}^{(\mathrm{old})}_8$. One possible explanation for this observation is that the Tanner graph of $\mathcal{C}^{(\mathrm{old})}_8$ contains subgraphs induced by the codeword of weight 6 while the minimum distance of $\mathcal{C}^{(\mathrm{new})}_8$ is 10. Allowing the Tanner graph of $\mathcal{C}_{\mbox{d}10}$ to have girth 6 but requiring that the minimum distance is at least 10 results in higher rate than the rate of $\mathcal{C}^{(\mathrm{new})}_8$, while maintaining the good error performance. This example clearly demonstrates that larger girth alone does not necessarily lead to better performance. We also remark that although minimum distance is used as the design
parameter to construct $\mathcal{C}_{\mbox{d}10}$ with good error performance, in general this may
not be sufficient to guarantee low error floors, since codes with high minimum distance may still contain trapping sets.
\begin{figure}
\begin{center}
\includegraphics[width = 3.45in]{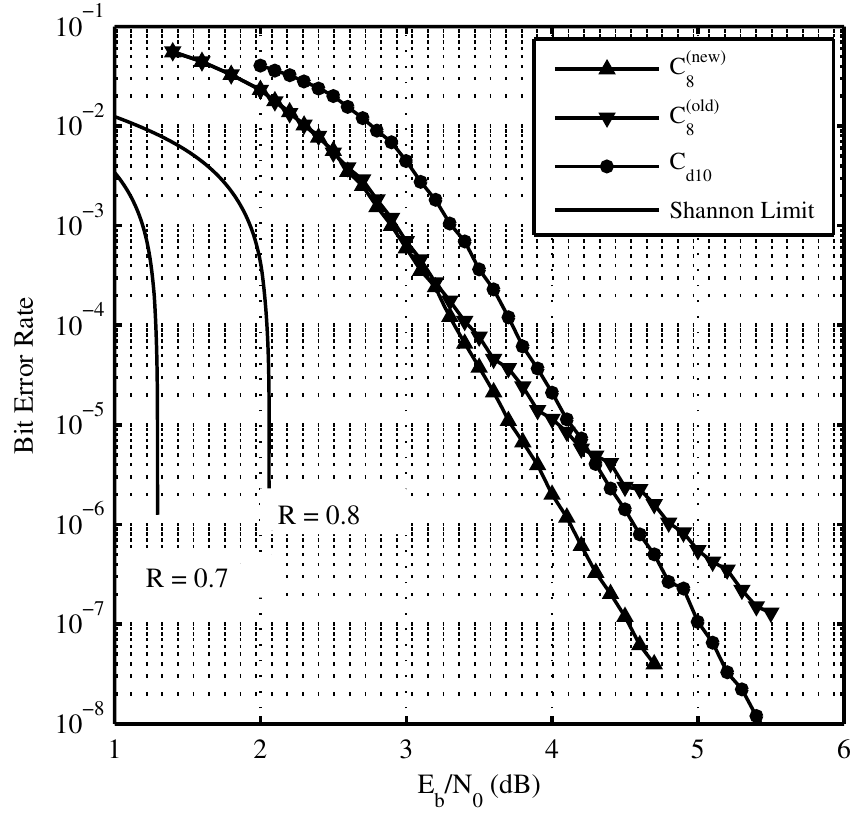}
\end{center}
\caption{Performances of the codes given in Example \ref{ex_530} over the AWGNC.}
\label{fg_503g8}
\end{figure}
\end{ex}


\begin{ex}\label{ex_demOne}
Let $q = 19^2 = 361$. The codes obtained when conditions $\tau_1$, $\tau_2$, $\tau_3$ and $\tau_4$ are imposed are denoted with $\mathcal{C}_1$, $\mathcal{C}_2$, $\mathcal{C}_3$ and $\mathcal{C}_4$, respectively. Conditions $\tau_1$, $\tau_2$, $\tau_3$ and $\tau_4$ are defined as
\begin{itemize}
\item $\tau_1$: $G$ has girth $g\geq 10$.
\item $\tau_2$: $G$ has girth $g\geq 8$; $G$ does not contain the (5,3) trapping set of girth 8 shown in Figure \ref{TS}\subref{532TS}; and $G$ does not contain the (6,4) trapping set shown in Figure \ref{TS}\subref{643TS}.
\item $\tau_3$: $G$ has girth $g\geq 8$; $G$ does not contain the (5,3) trapping set of girth 8 and an eight cycle in $G$ can share 2 variable nodes with at most one other eight cycle.
\item $\tau_4$: $G$ has girth $g\geq 6$; $G$ does not contain the (5,3) trapping set of girth 6 shown in Figure \ref{TS}\subref{531TS}; $G$ does not contain the (5,3) trapping set of girth 8; and an eight cycle in $G$ can share 2 variable nodes with at most one other eight cycle.
\end{itemize}

The Tanner graphs of these codes contain 361 check nodes. $\mathcal{C}_1$, $\mathcal{C}_2$, $\mathcal{C}_3$ and $\mathcal{C}_4$ have lengths $n_1=2888, n_2 = 3249, n_3=3610, n_4= 3971$ and rates $R_1 = 0.63, R_2 = 0.67, R_3 = 0.70, R_4 = 0.73$. The error performance of $\mathcal{C}_1$, $\mathcal{C}_2$, $\mathcal{C}_3$ and $\mathcal{C}_4$ under the SPA with a maximum of 50 iterations over the AWGNC is shown in Figure \ref{fg_DemOne}.

It can be seen that the conditions $\tau_1,\tau_2,\tau_3$ and $\tau_4$ are successively weaker. Since stronger conditions usually lead to codes with lower rates, we can observe in this example that $R_1<R_2<R_3<R_4$. From the simulation results, we see no loss in the error performance of codes with weaker conditions. This example emphasizes the importance of properly identifying the trapping sets to be avoided in the Tanner graph since it is crucial for the rate and the error performance of the code.
 
\textit{Remark:} Conditions $\tau_3$ and $\tau_4$ permit an eight cycle in $G$ to share 2 variable nodes with at most one other eight cycle. Consequently, ($6,4$) trapping sets can be present in the Tanner graphs but their variable nodes can be involved in at most two eight cycles. Therefore many children of the ($6,4$) trapping set are avoided (see \cite{ontology_vasic} for more details).
\begin{figure}
\begin{center}
\includegraphics{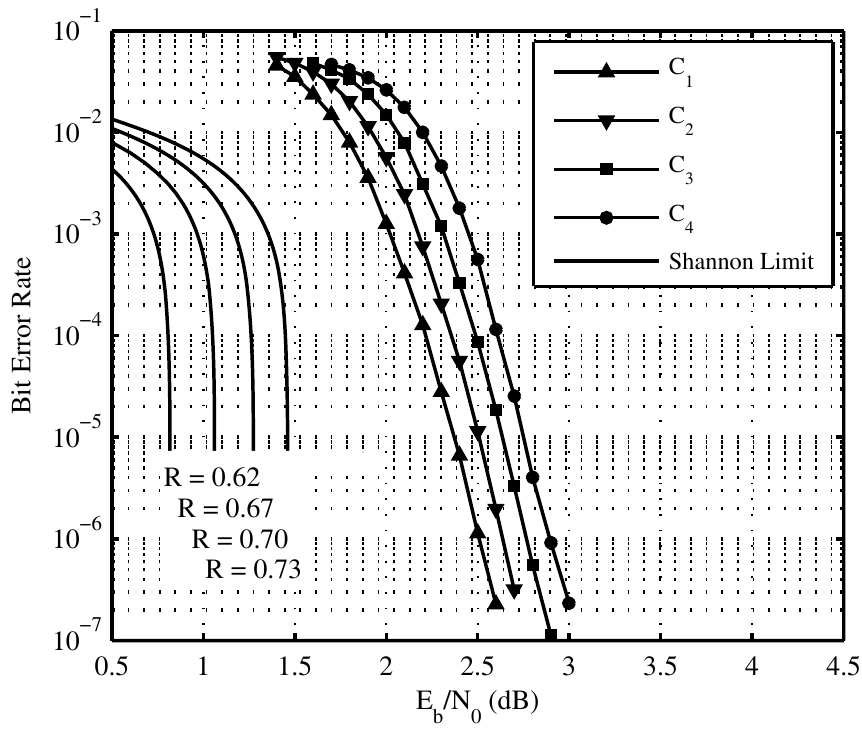}
\end{center}
\caption{Performances of the codes given in Example \ref{ex_demOne} over the AWGNC.}
\label{fg_DemOne}
\end{figure}
\end{ex}

%% file: Conclusion_v2.tex
\section{Discussion and Conclusions}\label{discussion}
We have introduced a class of structured LDPC codes with a wide range of rates and lengths. The code description is based on Latin squares, hence they can be explained both algebraically or combinatorially. Moreover, the description allows a code construction by progressively building the Tanner graph. The Tanner graph is built so that it does not contain a predefined set of trapping sets of iterative decoding algorithms. In this paper, we rely on the TSO - a database of trapping sets for the Gallager A/B algorithm on the BSC. Our conjecture is that trapping sets for other iterative decoding algorithms such as the SPA must contain trapping sets for the Gallager A/B algorithm. By eliminating trapping sets listed in the TSO, the codes have good error performance when decoded by other iterative decoding algorithms on the BSC or AWGNC. Although we could not provide enough experimental results for comparison with existing codes due to page limitations, to the best of our knowledge, our codes outperform the best known short length structured LDPC codes. Our current and future work includes identifying trapping sets for various decoding algorithms over the BSC and AWGNC, with the TSO as a starting point. 

%% file: Acknowledgment.tex
\section{Acknowledgment}
This work was funded by NSF under the grants IHCS-0725403, CCF-0634969, CCF-0830245.